\documentclass[12pt]{l4dc2021} 

\title[Input Convex Neural Networks for Building MPC]{Input Convex Neural Networks for Building MPC}
\usepackage{times}

\author{%
 \Name{Felix Bünning} \Email{felix.buenning@empa.ch}\\
 \Name{Adrian Schalbetter} \Email{adriasch@student.ethz.ch}\\
 \addr Automatic Control Laboratory, ETH Z{\"u}rich, Switzerland \\
 \addr Urban Energy Systems Laboratory, Empa D{\"u}bendorf, Switzerland\\
 \AND
 \Name{Ahmed Aboudonia} \Email{ahmedab@control.ee.ethz.ch}\\
 \Name{Mathias {Hudoba de Badyn}} \Email{mbadyn@ethz.ch}\\
 \addr Automatic Control Laboratory, ETH Z{\"u}rich, Switzerland \\
 \AND
 \Name{Philipp Heer} \Email{philipp.heer@empa.ch}\\
 \addr Urban Energy Systems Laboratory, Empa D{\"u}bendorf, Switzerland\\
 \AND
 \Name{John Lygeros} \Email{jlygeros@ethz.ch}\\
 \addr Automatic Control Laboratory, ETH Z{\"u}rich, Switzerland \\
}

\usepackage{gensymb}
\usepackage{tikz}
\usepackage{pgfplots}

\begin{document}

\maketitle

\begin{abstract}%
Model Predictive Control in buildings can significantly reduce their energy consumption. The cost and effort necessary for creating and maintaining first principle models for buildings make data-driven modelling an attractive alternative in this domain. In MPC the models form the basis for an optimization problem whose solution provides the control signals to be applied to the system. The fact that this optimization  problem has to be solved repeatedly in real-time implies restrictions on the learning architectures that can be used. Here, we adapt Input Convex Neural Networks that are generally only convex for one-step predictions, for use in building MPC. We introduce additional constraints to their structure and weights to achieve a convex input-output relationship for multi-step ahead predictions. We assess the consequences of the additional constraints for the model accuracy and test the models in a real-life MPC experiment in an apartment in Switzerland. In two five-day cooling experiments, MPC with Input Convex Neural Networks is able to keep room temperatures within comfort constraints while minimizing cooling energy consumption.
\end{abstract}

\begin{keywords}%
  Input Convex Neural Networks, Model Predictive Control, Building Energy Management%
\end{keywords}

\section{Introduction}

Model Predictive Control (MPC) in buildings can significantly reduce the energy consumption for space heating and cooling. However, developing and maintaining building models based on first principles is often considered expensive and tedious due to each building being individual \citep{Sturzenegger2016}. Here, data-driven modelling approaches are viewed as a promising alternative \citep{Bunning2020}. Ideally, such approaches should lead to accurate predictions for reliable controller performance and should be usable in convex optimization to find optimal control inputs in a limited amount of time.

As Artificial Neural Networks (ANN) have shown promising results in various domains \citep{Pimenidis2020, Silver2017}, they are natural candidates to be used as models in building MPC. However, a significant downside is that ANN generally do not lead to convex input-output mappings, making the resulting optimization problem intractable. To address this issue, \cite{Amos2017} present restrictions on the structure and weights of feed-forward ANN to build Fully Input Convex Neural Networks (FICNN), where the model output is convex with respect to all model inputs in single step predictions, and Partially Input Convex Neural Networks (PICNN), where the model output is convex with respect to a subset of the model inputs in single step predictions. They further demonstrate that such networks have high prediction accuracy in many domains, such as multi-label prediction, image completion, and reinforcement learning problems.
\cite{Chen2019} extend these formulations to a recurrent network structure for one-shot multi-step ahead predictions, which means that a sequence of outputs is predicted with a sequence of inputs in a single prediction. The authors apply the approach in an MPC scheme for MuJoCo locomotion tasks \citep{Todorov2012} and to building HVAC control in a simulation case in EnergyPlus.

In this work, we extend the work of \cite{Amos2017} to do multi-shot multi-step predictions with feed-forward networks; Predictions are made by re-evaluating the same network repeatedly, by using the outputs of previous timesteps as subsequent inputs. Here, the output at timestep $N$ is not only convex with respect to the input at timestep $N$, but also with respect to previous inputs. In comparison to \citep{Chen2019} this allows for more lightweight model architectures with less parameters to fit. We do this for both FICNN and PICNN by further constraining the network weights and by adding activation functions in appropriate places. We then compare the accuracy of the resulting networks to those presented by \cite{Amos2017} on a dataset from a real apartment. We also use the networks as a basis for an MPC controller in a real-life cooling experiment in the same apartment. Our experiments demonstrate the capability of the approach to keep room temperatures within comfort constraints while minimizing cooling energy consumption.

In Section \ref{sec: Methodlogy} we recap previously presented network structures and demonstrate their problem with keeping convexity in multi-step ahead predictions and MPC schemes. We then introduce further constraints on the networks and add activation functions to address this issue. We further show how the networks can be embedded in MPC schemes and address problems with lower state constraints. In Section \ref{sec: Case study} we first compare the prediction accuracy of our networks with previously presented ones. We then introduce the experimental case study and discuss the results. We conclude in Section \ref{sec: Conclusion}. 

\section{Methodology}
\label{sec: Methodlogy}

\subsection{Model Predictive Control}

MPC is a receding horizon optimization scheme for optimal control. At every time instant the state of the system $x_0$ is measured and an optimisation problem of the form

\begin{subequations}
\label{eqn:mpcmethod}
\begin{alignat}{2}
 \min_{\substack{u,x}}   &\quad && \sum_{k=0}^{N-1} J_k(x_{k+1},u_k)\label{eqn:mpctheory1}\\
\text{s.t. }   &\quad && \quad x_{k+1} = f(x_{k}, u_{k},d_k)      \label{eqn:mpctheory2}\\
 &\quad && \quad (x_{k+1},u_k) \in (\mathcal{X}_{k+1},\mathcal{U}_k)     \label{eqn:mpctheory3}\\
 &\quad && \forall k \in [0,...,N-1]     , \nonumber
\end{alignat}
\end{subequations}

\noindent is solved, where $x$, $u$ and $d$ denote states, inputs and disturbances respectively, $k$ denotes the timestep in the horizon $N$, $J$ denotes the cost function, $f$ denotes the system dynamics and $\mathcal{X}_k$, $\mathcal{U}_k$ encode desired constraints for states and inputs. The controller then applies the first element, $u_0^*$, of the optimal input sequence, $u^*$, to the system and the process is repeated.

To find optimal solutions to problem (1) fast and reliably, it is beneficial if the overall optimisation problem is convex in the decision variables $u$ and $x$. This implies restrictions on the cost function $J$, dynamics function $f$, and constraint functions $\mathcal{X}$ and $\mathcal{U}$. A common way for ensuring convexity is to select $J$ to be a convex function (typically, a convex quadratic), select $\mathcal{X}$ and $\mathcal{U}$ to be convex sets, and assume that f is a linear function. Here we show how convexity can be ensured for more general dynamics functions encoded by neural networks.

\subsection{Input Convex Neural Networks}

\cite{Amos2017} introduced an architecture for feed-forward Neural Networks where the scalar output of the network is convex with respect to all inputs (Fully Input Convex Neural Network or FICNN), or with respect to a subset of the inputs (Partially Input Convex Neural Network or PICNN). These networks are therefore promising candidates for $f$ in problem (1).

\begin{figure}
\centering
\begin{minipage}{.45\textwidth}
  \centering
		\includegraphics[width=0.9\textwidth]{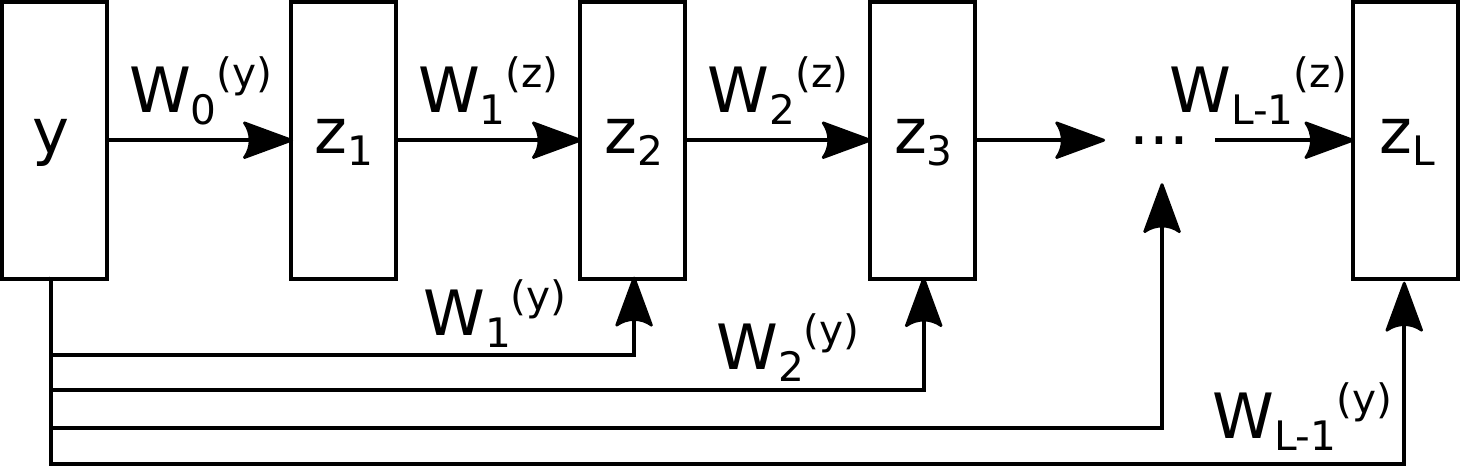}
	\caption{Schematic of a Fully Input Convex Neural Network}
	\label{fig: Schematic of a Fully Input Convex Neural Network}
\end{minipage}%
\begin{minipage}{0.1\textwidth}
\hspace{2mm}
\end{minipage}%
\begin{minipage}{.45\textwidth}
  \centering
		\includegraphics[width=0.9\textwidth]{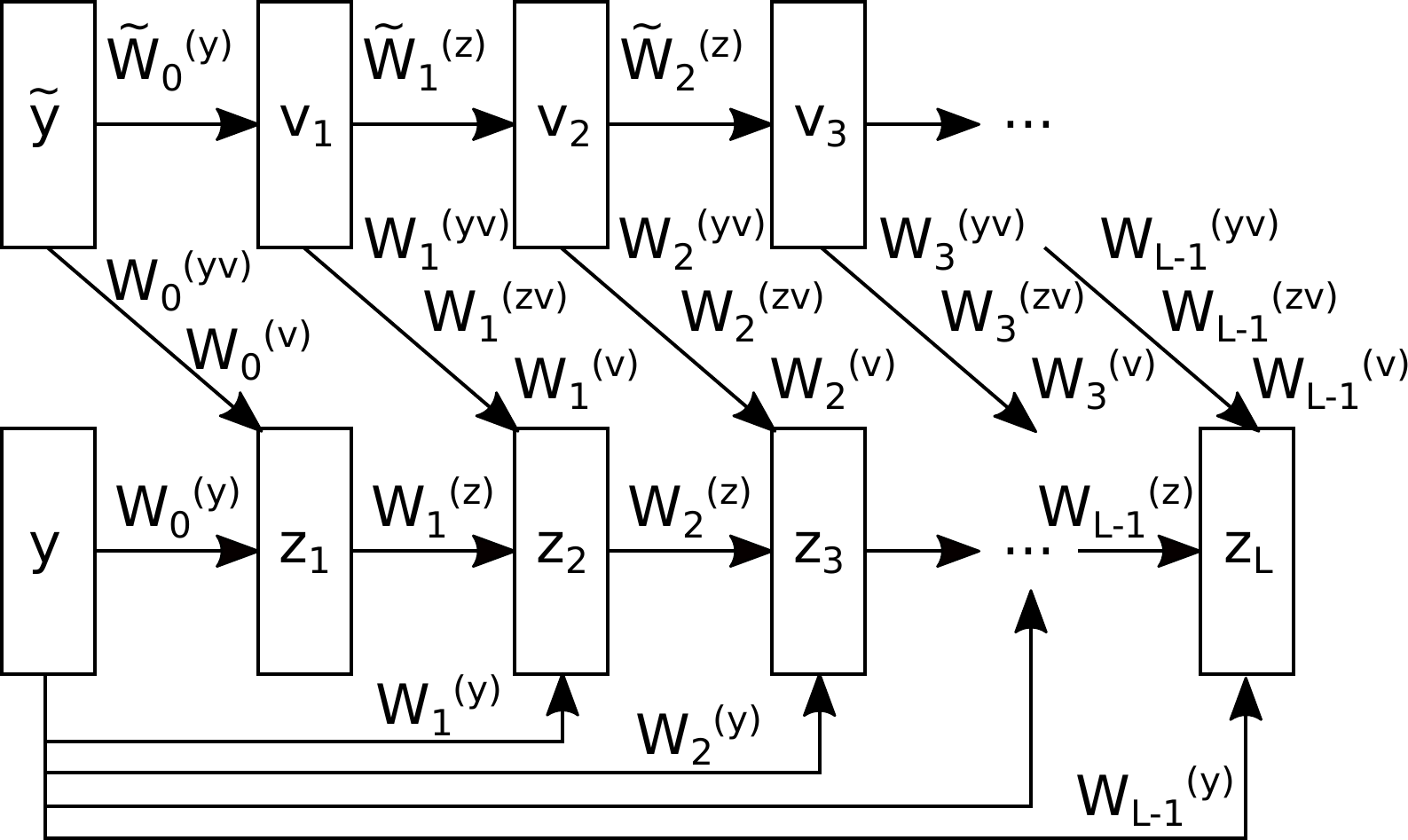}
	\caption{Schematic of a Partially Input Convex Neural Network}
	\label{fig: Schematic of a Partially Input Convex Neural Network}
\end{minipage}
\end{figure}

Figure \ref{fig: Schematic of a Fully Input Convex Neural Network} shows the schematic of an FICNN. It shows a $L$-layer fully connected network in which the output of each layer follows

\begin{equation}
z_{i+1}=g_i(W_i^{(z)}z_i+W_i^{(y)}y+b_i),
\label{eq: FICNN}
\end{equation}

\noindent with $z_0=0$ and $W_0^{(z)}=0$, where $g_i$ denotes the activation function for layer $i\in [1,..,L-1]$, $W_i^{(z)}$ and $W_i^{(y)}$ denote the network's weights, and $b_i$ denotes a constant bias. Both weights and biases are model parameters that are determined during network training. The network output $z_L$ is convex with respect to the input $y$ if all $W_{1:L-1}^{(z)}$ are non-negative and all activation functions $g_i$ are convex and non-decreasing \citep{Amos2017}.

Similarly, Figure \ref{fig: Schematic of a Partially Input Convex Neural Network} depicts a PICNN. Here, the output of each network layer follows

\begin{equation}
\label{eqn:picnnmethod}
\begin{aligned}
  v_{i+1} &= \Tilde{g}_i(\Tilde{W}_i v_i + \Tilde{b}_i), \\
  z_{i+1} &= g_i \bigg[W_i^{(z)} \Big\{ z_i \circ [W_i^{(zv)} v_i + b_i^{(z)}] \Big\} + 
  &W_i^{(y)} \Big\{ y \circ [W_i^{(yv)} v_i + b_i^{(y)}] \Big\} + W_i^{(v)} v_i + b_i\bigg],  \\
\end{aligned}
\end{equation}

\noindent with $z_0, W_0^{(z)}=0$ and $v_0=\tilde{y}$, where $\tilde{g}_i$ and $g_i$ are activation functions, $\tilde{W_i}$, $W_i^{(z)}$, $W_i^{(zv)}$, $W_i^{(yv)}$, $W_i^{(v)}$ are input weights, $\Tilde{b}_i$, $b_i$, $b_i^{(z)}$, $b_i^{(y)}$ are constant biases, and $\circ$ denotes the Hadamard product. Under the condition that all weights $W_{1:L-1}^{(z)}$ are non-negative and all activation functions $g_i$ are convex and non-decreasing the model output $z_L$ is convex with respect to $y$ (but not necessarily with respect to $\tilde{y}$).  \citep{Amos2017}

FICNN and PICNN guarantee input-output convex behaviour, making them promising candidates for MPC schemes, as mentioned above. However, when ICNNs are applied for $f$ in this context, for example as $x_{k+1} = f(y=(x_{k}, u_{k},d_k))$ in problem (1), it becomes evident that there is an issue with convexity in multi-step ahead prediction. While the first step $x_{1}=f(x_0,u_0,d_0)$ is convex, the second one is not guaranteed to be. For example, if a 1-layer FICNN is considered, the state at timestep 2 follows

\begin{equation}
x_2=f(x_1,u_1,d_1)=g_0(W_0^{(y)}(x_1, u_1, d_1)+b_0).
\label{eq: x2 example}
\end{equation}

\noindent As $x_1$ is a convex function of $x_0$ and $u_0$, for example $x_1=x_0^2+u_0^2$, and $W_0^{(y)}$ can attain any value (thus also a negative one), the term $W_0^{(y)}(x_1,u_1,d_1)$ can become concave in $x_0,u_0$, for example $(-1)(x_0^2+u_0^2)$. The state $x_2$ is thus not guaranteed to be convex with respect to $x_0$ and $u_0$.

\subsection{ICNN for multi-step ahead prediction}
\label{sec: ICNN for multi-step ahead prediction}

To make ANNs input convex in the face of multi-step ahead prediction, \cite{Chen2019} have presented a solution for Fully Input Convex Recurrent Neural Networks. Here, we present a solution for Fully and Partially Input Convex Feed-Forward Neural Networks. We accomplish this through additional constraints in the network architecture and the use of ReLu activation functions in appropriate places in the network topology. Our approach readily extends to FICNN, where it becomes the feed-forward counterpart of the approach developed by \cite{Chen2019} for recurrent neural networks.

\vspace{5mm}

\noindent \textbf{Proposition 1:} \textit{Consider the two FICNNs $f_1(y_1)$ and $f_2(y_2)$ as defined in eq. \eqref{eq: FICNN}. The composition $f_2(y_2,f_1(y_1))$ is convex with respect to $y_1$, if all weights $W_i^{(z)}$ and $W_i^{(y)}$ are non-negative and all functions $g_i$ are convex and non-decreasing.}\\

The proof follows from the fact that non-negative sums of convex functions are convex and that compositions of a convex function and a convex non-decreasing function are also convex. Looking at eq. \eqref{eq: FICNN}, $W_i^{(z)}z_i$ is convex assuming that $z_i$ is convex and $W_i^{(z)}$ is non-negative. The parameter $b_i$ is a constant. The term $W_i^{(y)}y$ is convex in $y$ for any $W_i^{(y)}$ if $y$ is constant because a linear function with negative gradient is also convex \cite{Amos2017}. Here, we require $W_i^{(y)}$ to be non-negative, because $y$ itself might be a convex function in the form of an ICNN. The term $(W_i^{(z)}z_i+W_i^{(y)}y+b_i)$ is therefore a non-negative sum of convex functions and $g_i(W_i^{(z)}z_i+W_i^{(y)}y+b_i)$ a composition of a convex function and a convex non-decreasing function (for example the commonly used ReLu function $g_i=max(x,0)$).
For the example of eq. \eqref{eq: x2 example}, as $W_0^{(y)}$ is now constrained to be non-negative, $W_0^{(y)}(x_1=f(x_0,u_0,d_0))$ is convex in $x_0,u_0$, if $x_1$ is convex in $x_0,u_0$ (which is the case). Thus, $x_2$ is also convex in $x_0$ and $u_0$. The network output $z_L$ is a convex non-decreasing function of the input $y$.

\vspace{5mm}

In the case of PICNN, we propose the following structure for the outputs of the network layers for input convex multi-step prediction,

\begin{equation}
\label{eqn:picnnmethodology}
\begin{aligned}
  v_{i+1} &= \Tilde{g}_i(\Tilde{W}_i v_i + \Tilde{b}_i) \\
  z_{i+1} &= g_i \bigg[W_i^{(z)} \Big( z_i \circ g_i^{(zv)}[W_i^{(zv)} v_i + b_i^{(z)}] \Big) + 
  &W_i^{(y)} \Big( y \circ g_i^{(yv)}[W_i^{(yv)} v_i + b_i^{(y)}] \Big) + W_i^{(v)} v_i + b_i\bigg],
\end{aligned}
\end{equation}

\noindent where the activation functions $g_i^{(zv)}$ and $g_i^{(yv)}$ are added compared to eq. \eqref{eqn:picnnmethod}.

\vspace{5mm}

\textbf{Proposition 2:} \textit{Consider the two PICNNs $f_1(\tilde{y}_1, y_1)$ and $f_2(\tilde{y}_2, y_2)$ as defined in eq. \eqref{eqn:picnnmethodology}. The composition $f_2(\tilde{y}_2,y_2,f_1(\tilde{y}_1,y_1))$ is convex with respect to $y_1$, if all weights $W_i^{(z)}$ and $W_i^{(y)}$ are non-negative, all functions $g_i^{(zv)}$ and $g_i^{(yv)}$ map to a non-negative value and the function $g_i$ is convex and non-decreasing.}\\


\noindent The proof again follows from only applying operations that maintain convexity. Going through eq. (\ref{eqn:picnnmethodology}) term by term, $z_i \circ g_i^{(zv)}[W_i^{(zv)} v_i + b_i^{(z)}]$ is convex in inputs of $z_i$ if $z_i$ is a convex function because $g_i^{(zv)}$ maps all negative values of $[W_i^{(zv)} v_i + b_i^{(z)}]$ to zero. As $W_i^{(z)}$ is non-negative, $W_i^{(z)} ( z_i \circ g_i^{(zv)}[W_i^{(zv)} v_i + b_i^{(z)}] )$ is also convex. The same argument can be made for the next term $W_i^{(y)} ( y \circ g_i^{(yv)}[W_i^{(yv)} v_i + b_i^{(y)}])$ regarding the convexity in $y$ and inputs of $y$. Here, we need $g_i^{(yv)}$ to map all negative values to zero, because $y$ is not a constant (as in \cite{Amos2017}), but a convex function itself. As $g_i$ is convex non-decreasing, the composition of $g_i$ and before-mentioned terms is convex. Eq. (\ref{eqn:picnnmethodology}) is thus convex in any convex $y$ and $z_i$. We note in passing, that other formulations of FICNN and PICNN can be thought of and will give the same result as long as it is ensured that $z_{i+1}$ is convex non-decreasing in $y$.

%

\subsection{Embedding ICNN in (quasi-)convex MPC}

The adaptations in Section \ref{sec: ICNN for multi-step ahead prediction} can be used to formulate a convex MPC scheme \citep{Boyd2004} commonly used for building control as

\begin{subequations}
\label{eqn:mpcmethod2}
\begin{alignat}{2}
 \min_{\substack{u,x}}   &\quad && \sum_{k=0}^{N-1} J_k(x_{k+1},u_k)\label{eqn:mpctheory1}\\
\text{s.t. }   &\quad && \quad x_{min} \leq \Big( x_{k+1} = f ( x_k , u_{k},d_k ) \Big) \leq x_{max} \label{eqn:mpctheory2}\\
 &\quad && \quad u_{min}  \leq u_k \leq u_{max}     \label{eqn:mpctheory3}\\
 &\quad && \forall k \in [0,...,N-1]     , \nonumber
\end{alignat}
\end{subequations}

\noindent where $x$, $u$ and $d$ are states, inputs and disturbances respectively. $J$ is an appropriate convex cost function, $f$ are the dynamics represented by an ICNN, $u_{min}$ and $u_{max}$ are lower and upper input constraints and $x_{min}$ and $x_{max}$ are lower and upper state constraint. Note, that lower state constraints are generally not possible for $f$ being convex, as super-level sets of convex functions are generally not convex. However, in the case of using the presented ICNN, $f$ is convex and non-decreasing, which means that also super-level sets are convex. We can thus add a lower state constraint $x_{min} \leq x_{k+1}$ in our approach.

\begin{figure}[h]

\centering
\begin{minipage}{.48\textwidth}
    \centering
    \begin{tikzpicture}
        \begin{axis}[axis x line=center, width=5cm,height=4cm,
                      axis y line=center,
                      xlabel={$u$},
                      xtick={0.5,1.2321},
                      xticklabels={},
                      ytick={2.0,4.0},
                      yticklabels={$x_{min}$,$x_{max}$},
                      ylabel={$x(u)$},
                      xlabel style={below},
                      ylabel style={above left},
                      xmin=0,
                      xmax=1.7,
                      ymin=-0,
                      ymax=5.5]
            \addplot[line width = 0.35mm, red, dashed, domain=-2:0.5] {-(x+0.5)^2+3} node [pos=1.0, above] {${\epsilon}^*$};  
            \addplot[line width = 0.35mm, black, domain=-2:2] {(x+0.5)^2+1};
            \addplot[line width = 0.35mm, red, domain=1.2321:2] {(x+0.5)^2+1} node [pos=0.1,label={${\epsilon}^*$}] {};  
            \addplot[line width = 0.35mm, red, domain=0:0.5] {(x+0.5)^2+1};           
            \addplot[line width = 0.35mm, black, dotted, domain=-2.5:2.5] {2.0} node [pos=0.5, below right]{};
            \addplot[line width = 0.35mm, black, dotted, domain=-2.5:2.5] {4.0} node [pos=0.5, below right]{};
   
        \end{axis}
    \end{tikzpicture}
    \caption{Optimal slack variable ${\epsilon}^*$ at upper and lower state constraints \newline }
    \label{fig: slack 1}
\end{minipage}
\begin{minipage}{0.02\textwidth}
\hspace{2mm}
\end{minipage}%
\begin{minipage}{.48\textwidth}
    \centering
    \begin{tikzpicture}
        \begin{axis}[axis x line=center, width=5cm,height=4cm,
                      axis y line=center,
                      xlabel={$u$},
                      xtick={0.5,1.2321},
                      xticklabels={},
                      ytick={-4},
                      yticklabels={},
                      ylabel={$J({\epsilon}^*(u))$},
                      xlabel style={below},
                      ylabel style={above right},
                      xmin=0,
                      xmax=1.7,
                      ymin=-0,
                      ymax=5.5]
            \addplot[line width = 0.35mm, red, domain=-2:0.5] {-(x+0.5)^2+3} node [pos=1.0, above ] {${\epsilon}^*$};  
            \addplot[line width = 0.35mm, black, domain=0.5:1.2321] {2};
            \addplot[line width = 0.35mm, red, domain=1.2321:2] {(x+0.5)^2-1} node [pos=0.1,label={${\epsilon}^*$}] {};

        \end{axis}
    \end{tikzpicture}
    \caption{Quasiconvex cost function $J$ as a result of concave non ascending slack variable at lower state constraint}
    \label{fig: slack 2}
    
\end{minipage}
\end{figure}

Soft state constraints in the form of slack variables are often necessary in practical applications of MPC to ensure feasibility at all times. Adding soft constraints changes the problem to

\begin{subequations}
\label{eqn:mpcmethod2}
\begin{alignat}{2}
 \min_{\substack{u,x,\epsilon}}   &\quad && \sum_{k=0}^{N-1} J_k(x_{k+1},u_k, {\epsilon}_k)\label{eqn:mpctheory1}\\
\text{s.t. } &\quad && \quad x_{min} - {\epsilon}_k \leq f ( x_k , u_{k},d_k ) \leq x_{max} + {\epsilon}_k  \label{eqn:mpctheory2}\\
 &\quad && \quad {\epsilon}_k \geq 0     \label{eqn:mpctheory3}\\
 &\quad && \quad u_{min} \leq u_k \leq u_{max}    \label{eqn:mpctheory3}\\
 &\quad && \forall k \in [0,...,N-1]     , \nonumber
\end{alignat}
\end{subequations}

\noindent where $\epsilon$ is the slack variable, and $J$ is now a function of $x$, $u$ and $\epsilon$. However, problem (7) is not convex any more. While ${\epsilon}^*$ (i.e the $\epsilon$ that minimizes $J$) is a convex non-decreasing function of $u$ for $x \geq x_{max}$, as depicted on the right of Fig. \ref{fig: slack 1}, it is a concave non-ascending function for $x \leq x_{min}$, as depicted on the left of Fig. \ref{fig: slack 1}. $J$ is therefore not convex at the lower state constraint (for $x \leq x_{min}$). We note however, that $J$ is quasiconvex \citep{Boyd2004}, as shown for the example $J= const + \epsilon$ in Fig. \ref{fig: slack 2}, and can thus be solved to global optimum with many solvers.

\section{Case studies}
\label{sec: Case study}

\subsection{Set up}

We apply both introduced networks in a numerical and an experimental case study in the Urban Mining and Recycling (UMAR) unit of the NEST demonstrator building \citep{Richner2018}. The unit is an occupied apartment comprising two bedrooms, a living room, two bathrooms, and an entrance area. The rooms are equipped with ceiling heating and cooling panels, which are connected to the central heating and cooling system of NEST through a heat exchanger. In standard operation, the room temperature is controlled through thermostats that open or close valves to the ceiling panels. The supply temperature and the supply pump pressure are constant.

In different case studies, the ICNN are used to predict the room temperature $T_{br,k+1}=T_{br,k} + \Delta T_{br, k+1}$ of one of the bedrooms, by predicting the temperature change

\begin{equation}
\begin{aligned}
\Delta T_{br, k+1} = f \Big( & \dot{Q}_{sol,k},\dot{Q}_{sol,k-1},\dot{Q}_{sol,k-2}, t_{sin}, t_{cos},\\ & \delta T_{l,k}, \delta T_{amb,k}, \Delta T_{br,k},\Delta T_{br,k-1}, \Delta T_{br,k-2}, Q_{u,k} \Big),
\end{aligned}
\end{equation}

\noindent where, $\dot{Q}_{sol,k},\dot{Q}_{sol,k-1},\dot{Q}_{sol,k-2}$ are the global solar irradiation at the current timestep and the two previous timesteps, $t_{sin}$ and $t_{cos}$ are the time of the day encoded as a sine and cosine function, $\delta T_{l,k}$ and $\delta T_{amb,k}$ are the temperature differences between the bedroom and living room, and the bedroom and ambient at timestep $k$ respectively, and $Q_{u,k}$ denotes the heating/cooling energy (i.e. the control input). Note, that $\Delta$ defines differences in terms of time, while $\delta$ defines differences in terms of location. The feature selection is a result of an extensive study based on k-fold cross validation, where measurement data of 1 year from the UMAR unit is used \citep{Schalbetter2020}. Function $f$ represents either a FICNN or a PICNN; in the case of PICNN, the features are divided into convex features $y=(\delta T_{amb,k}, \Delta T_{br,k},\Delta T_{br,k-1},$ $ \Delta T_{br,k-2}, Q_{u,k})$, which are related to the state or control inputs, and features $\tilde{y}=(\dot{Q}_{sol,k},\dot{Q}_{sol,k-1},\dot{Q}_{sol,k-2},t,\delta T_{l,k})$, which do not require convexity because they are related to the disturbances and are not optimized over.\footnote{Note, that $\delta T_{l,k}$ is kept constant during the prediction and thus does not need to be in the convex inputs.}

We use models with two different sampling rates, 20 minutes and 180 minutes, for predictions up to 1 hour and predictions $>$1 hour respectively, as models with larger sampling rates showed better prediction performance for long horizons in preliminary experiments. The hyperparameters for both models can be seen in Table \ref{table:hyperparams}. All parameter decisions are results of the cross validation study \citep{Schalbetter2020}. To allow the network output $z_L$ (i.e. $\Delta T_{br, k+1}$) to be negative, we use a shifted ReLu function, $g(x)=max(x,0)-\beta$, as the activation function in the output layer instead of a regular one. Here, $\beta$ is a hyperparameter of the network.

\tiny

\begin{table}[h]
\scriptsize
\begin{center}
\bgroup
\def\arraystretch{1.5}
\begin{tabular}{ l c c }
\hline
 \textbf{hyperparameters} & \textbf{1 hour model} & \textbf{$>$1 hour model}\\
\hline
 optimizer & Adam & Adam\\
\hline
 epochs & 20 & 40\\
\hline
 layers & $4$ & $4$\\
\hline
 nodes per layer & 9 & 8\\    
\hline
 ReLU-offset ($\beta$) & 0.8 & 12\\
\hline
\end{tabular}
\egroup
\end{center}
\caption{Tuned hyperparameters for the ICNN for 1 hour (left) and $>$1 hour predictions (right).}
\label{table:hyperparams}
\end{table}

\normalsize

\subsection{Numerical case study}

In the numerical study, we compare the prediction performance of our ICNN to those introduced by \cite{Amos2017} for room temperature predictions of 1 hour and 6 hours. To account for the fact that training neural networks is a non-convex problem, we divide the measurement data from UMAR into 12 folds of size one month, use 9 randomly selected folds for training, the remaining 3 folds for validation and repeat this step 100 times for each network.

\begin{figure}
  \centering
		\includegraphics[width=1.0\textwidth , trim={1cm 0 1cm 0}]{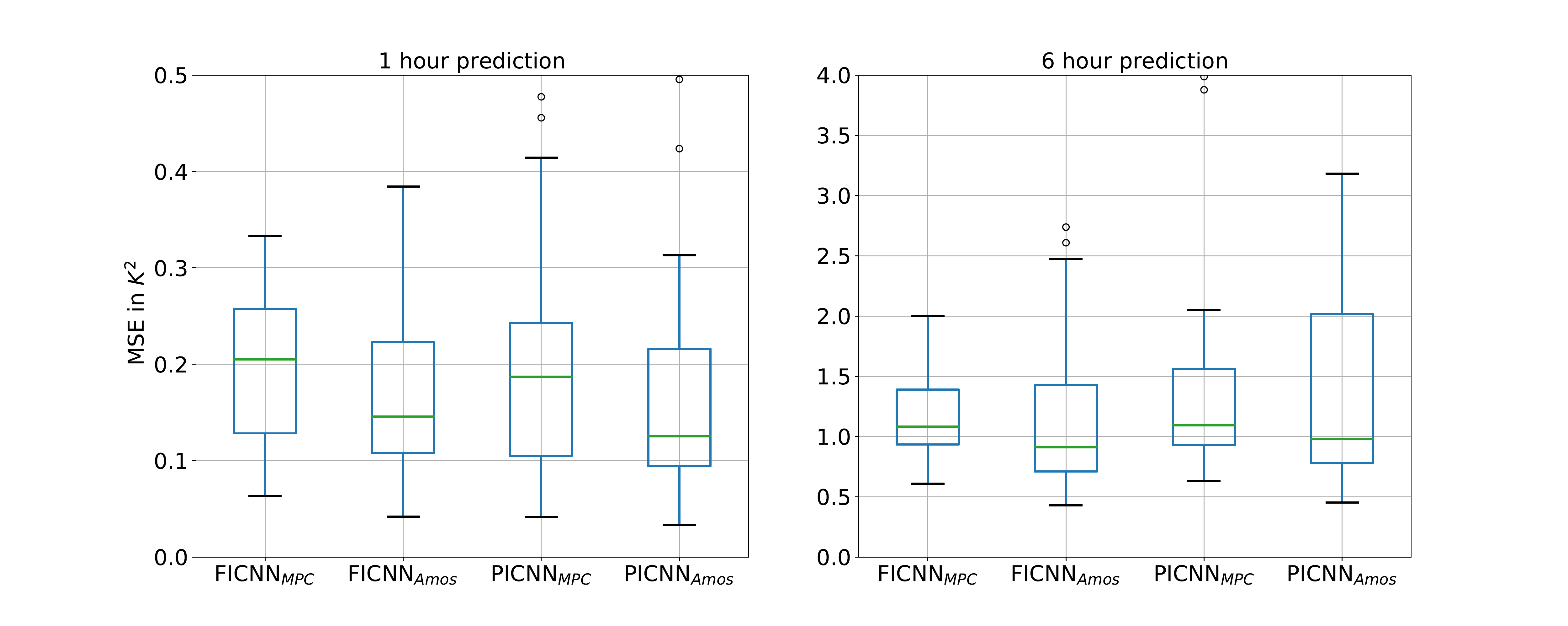}
	\caption{Prediction accuracy of different ICNN for 1 hour predictions and 6 hour predictions in UMAR}
	\label{fig: Prediction accuracy of different ICNN for 1 hour predictions and 6 hour predictions in UMAR}
\end{figure}

The results of the comparison are shown in Figure \ref{fig: Prediction accuracy of different ICNN for 1 hour predictions and 6 hour predictions in UMAR}. The left plot shows the boxplots of the mean squared error (MSE) in $K^2$ for 1 hour predictions, with median, interquatile range (IQR), and minimum and maximum for the different types of networks. The index $MPC$ denotes the networks presented in this article, while the index $Amos$ denotes the ones presented in \cite{Amos2017}. It can be seen that the median of the MSE is significantly lower for the Amos networks, for both FICNN and PICNN. The IQR is comparable for all networks, while minimum and maximum do not show a decisive trend. In the case of the 6 hour prediction, depicted on the right of Figure \ref{fig: Prediction accuracy of different ICNN for 1 hour predictions and 6 hour predictions in UMAR}, the IQR and min-max range are smaller for our networks, while the median is again lower for the networks presented in \cite{Amos2017}. The increase in median compared to \cite{Amos2017} can be expected for general system dynamics, as our networks are restricted to convex non-decreasing functions while the those of \cite{Amos2017} allow more general convex functions. However, as buildings are generally positive \citep{Khosravi2019} and monotonous systems, for example the room temperature is a positive monotonous function of the valve position of a radiator or of the supply temperature of the heating system, the unambiguity of the result is nevertheless surprising. A possible explanation is that removing the monotonicity requirement allows a better approximation of internal gains from occupancy as a function of the time and solar gains through windows as a function of the time and the global horizontal irradiation.

\subsection{Experimental case study}

In two real-life experiments we have applied the ICNN to MPC for room temperature control in a bedroom of the UMAR apartment during the cooling season. The MPC set up corresponds to problem (8) with the cost function $J_k=R u_k^2+\lambda {\epsilon}_k^2$. The problem parameters are set to $R=1$, $\lambda=100$, which proved to be successful in earlier studies \citep{Bunning2020}. There are no state costs, which is a common choice, as there are usually no stability issues with buildings. The horizon is set to 7 hours, divided into three predictions with the network with 20 minutes sampling time and two with 180 minutes sampling time. \textit{Move blocking} \citep{Cagienard2007} is applied here because longer sampling times led to better prediction accuracy for long horizons in preliminary studies. The input constraints are $u_{min}=-0.6$ kWh and $u_{max}=0$ kWh and are implemented with pulse-width modulation of the supply valves. The state constraints $x_{min}$ and $x_{max}$ reflect comfort constraints for the room temperatures and are time varying and shown in the results. The scheme was programmed in Python 3 and COBYLA \citep{Powell1994} was used as a solver. It is solved in less than ten seconds on a laptop computer.

\begin{figure}
  \centering
		\includegraphics[width=1.0\textwidth , trim=0 0.5cm 0 0]{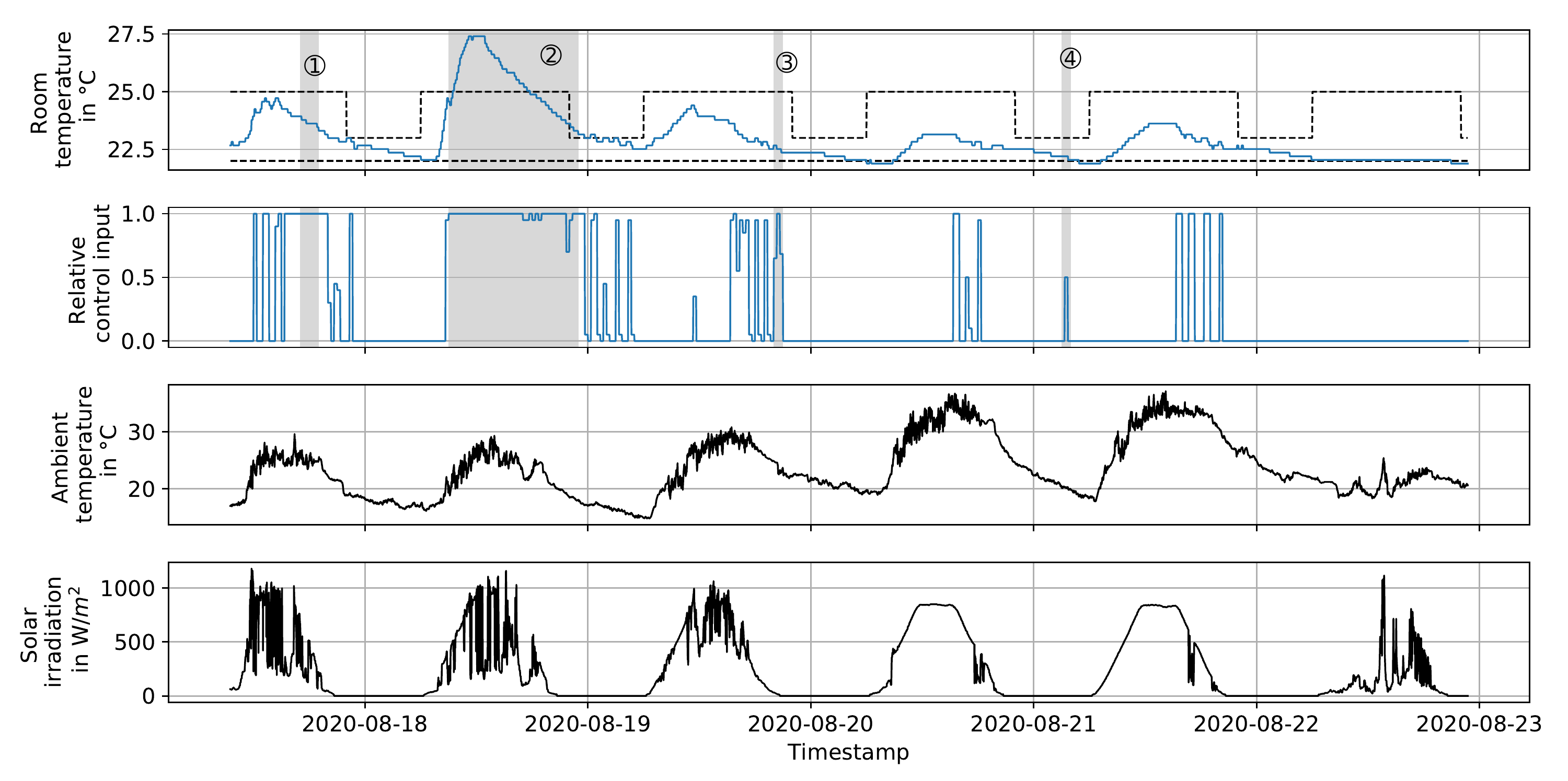}
	\caption{Experimental results of room temperature control by MPC with a FICNN}
	\label{fig: Exp1}
\end{figure}

\begin{figure}
  \centering
		\includegraphics[width=1.0\textwidth , trim=0 0.5cm 0 0]{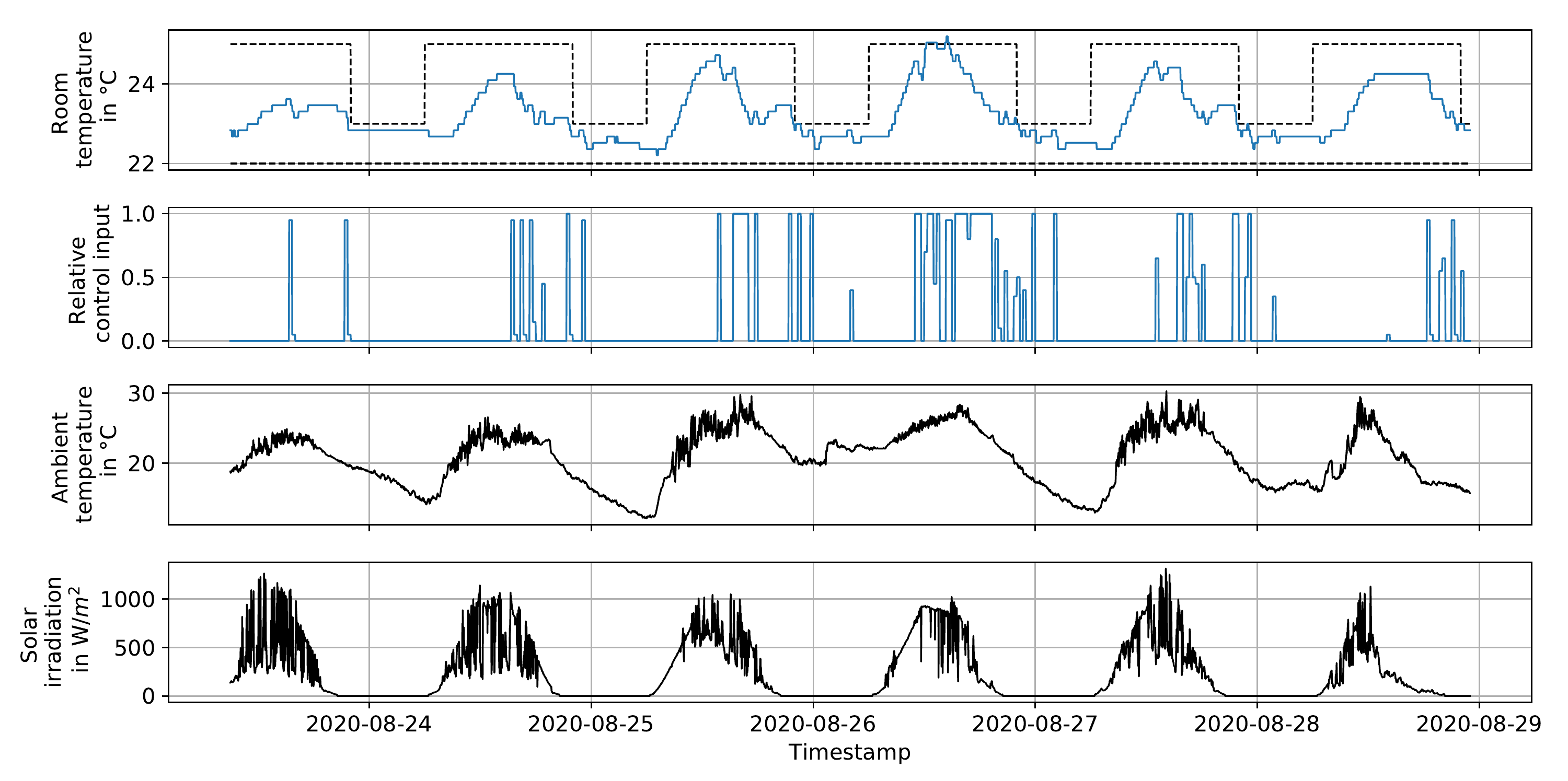}
	\caption{Experimental results of room temperature control by MPC with a PICNN}
	\label{fig: Exp2}
\end{figure}

Figure \ref{fig: Exp1} shows the results of MPC in combination with a FICNN applied to one of the bedrooms of UMAR. The first plot depicts the room temperature in blue and the time-varying comfort constraints in dashed black. The upper comfort constraint is 25 \degree C between 6 am and 10 pm, and 23 \degree C otherwise. The lower comfort constraint is constant at 22 \degree C. The second plot shows the relative control (cooling) input. It can be seen that the MPC controller keeps the temperature between the comfort constraints during most times and allows the temperature to rise during times when the upper comfort constraint is higher. One exception is the the second day (marked with \raisebox{.5pt}{\textcircled{\raisebox{-.9pt} {2}}}). Here, the normally shut window blinds were automatically opened due to high winds, and the cooling system was not able to compensate the solar gains although running at full capacity. During the other times that are marked in grey (and with {\textcircled{\raisebox{-.9pt} {1}}}, {\textcircled{\raisebox{-.9pt} {3}}}, {\textcircled{\raisebox{-.9pt} {4}}}), connection to the actuators was lost and the standard thermostat controller took over. The lower two plots show the ambient conditions, which are measured at the top of the building.

Figure \ref{fig: Exp2} shows the results of the second experiment, where MPC with a PICNN was applied in the other bedroom. As this bedroom is more sensitive to solar gains due to a neighbouring apartment, the temperature rise during the day is much more visible. It is evident that the used PICNN has sufficient prediction accuracy for MPC because the controller exactly meets the lowered upper comfort constraint at 10 pm on every single occasion. Comparing this result to the performance of FICNN in Figure \ref{fig: Exp1} it appears that the additional generality of PICNN pays off, as the FICNN tends to cool down the room more than necessary. As the experiments are conducted in different rooms on different days, the evidence is not conclusive, but there is at least a strong indication in this direction.

\section{Conclusion}
\label{sec: Conclusion}

In this work, we have proposed constraints and activation functions to make ICNN input-convex, not only for single step predictions, but also for multi-step ahead predictions. Although these networks show a decreased model accuracy compared to one-step ahead ICNN, the adoptions enable the networks to be used in (quasi-)convex MPC schemes. In two real-life experiments with ICNN in building energy MPC, the controller kept the room temperatures within comfort constraints, while exploiting time periods with relaxed constraints to save cooling energy. Ongoing research focuses on more general formulations of ICNN and and constraints in related MPC formulations.

\acks{This research project is financially supported by the Swiss Innovation Agency Innosuisse and is part of the Swiss Competence Center for Energy Research SCCER FEEB{\&}D.}

\bibliography{MyCollection}

\end{document}